\documentclass[12pt, a4paper, twoside]{scrartcl}

\pdfoutput=1

\usepackage[tracking=true,expansion=true,stretch=15,shrink=15]{microtype}
\DeclareMicrotypeSet*[tracking]{my}
{ font = */*/*/sc/* }
\SetTracking{encoding = *, shape = sc }{ 45 }
\SetProtrusion{encoding={*},family={bch},series={*},size={6,7}}
{1={ ,750},2={ ,500},3={ ,500},4={ ,500},5={ ,500},
	6={ ,500},7={ ,600},8={ ,500},9={ ,500},0={ ,500}}
\SetExtraKerning[unit=space]
{encoding={*}, family={qhv}, series={b}, size={large,Large}}
{1={-200,-200},
	\textendash={400,400}}

\usepackage{setspace}
\usepackage[
	left=1.3in,
	right=1.3in,
	top=0.7in,
	bottom=0.7in,
	includeheadfoot
]{geometry}
\usepackage[automark, headsepline]{scrpage2}

\usepackage[disable]{todonotes}
\usepackage{lineno}
\usepackage{amssymb}
\usepackage{amsmath}
\usepackage{amsthm}
\usepackage{dsfont}
\usepackage{bm}
\usepackage{enumerate}
\usepackage{graphicx}
\usepackage{subfigure}
\usepackage{float}
\usepackage{tikz}
\usetikzlibrary{shapes,arrows,decorations.pathmorphing,graphs,positioning,backgrounds}
\usepackage[round,comma]{natbib}
\usepackage{hyperref}
\usepackage{nameref}
\definecolor{TUMblue}{cmyk}{1, .54, .04, .19}
\hypersetup{
	colorlinks   = true, 
	urlcolor     = TUMblue, 
	linkcolor    = TUMblue, 
	citecolor   = TUMblue 
}

\newtheoremstyle{new}{12pt}{12pt}{\itshape}{}{\bfseries}{.}{1em}{}
\theoremstyle{new}

\allowdisplaybreaks
\begin{document}

\title{Vine copula based post-processing of ensemble forecasts for temperature}

\author{A. M\"oller\footnote{Corresponding author, Institute of Applied Stochastics and Operations Research, Clausthal University of Technology, Clausthal-Zellerfeld, Germany, (email:\href{mailto:annette.moeller@tu-clausthal.de}{annette.moeller@tu-clausthal.de})}, L. Spazzini\footnote{Partners4Innovation, Milano, Italy}, D. Kraus\footnote{Bayrische Landesbank, Munich, Germany}, T. Nagler, C. Czado \footnote{Center of Mathematics, Technical University of Munich, Garching, Germany}}

\date{\hspace{3pt} \normalsize\today}

\maketitle


\begin{abstract}
\noindent {\bfseries \sffamily Abstract}\\
Today weather forecasting is conducted using numerical weather prediction (NWP) models, consisting of a set of differential equations describing the dynamics of the atmosphere. The output of such NWP models are single deterministic forecasts of future atmospheric states.
To assess uncertainty in NWP forecasts so-called forecast ensembles are utilized. They are generated by employing a NWP model for distinct variants.
However, as forecast ensembles are not able to capture the full amount of uncertainty in an NWP model, they often exhibit biases and dispersion errors. Therefore it has become common practise to employ statistical post processing models which
correct for biases and improve calibration.
We propose a novel post processing approach based on D-vine copulas, representing the predictive distribution by its quantiles. These models allow for much more general dependence structures than the
state-of-the-art EMOS model and is highly data adapted.
Our D-vine quantile regression approach shows excellent predictive performance in comparative studies of temperature forecasts over Europe with different forecast horizons based on the 52-member ensemble of the European Centre for Medium-Range Weather Forecasting (ECMWF). Specifically for larger forecast horizons the method clearly improves over the benchmark EMOS model.\\[12pt]
{\itshape Keywords: Probabilistic Forecasting; Ensemble post processing; Ensemble model output statistics; D-vine copulas; calibration of NWP forecasts.}
\end{abstract}


\pagestyle{scrheadings}
\clearscrheadings
\lohead{D-vine based post-processing}
\rohead{\pagemark}
\lehead{M\"oller, Spazzini, Kraus, Nagler, Czado}
\rehead{\pagemark}

\section{Introduction} \label{sec:introduction}

The main sources of uncertainty in numerical weather prediction (NWP)
lie in the choice of a physical model and its boundary conditions. These are typically addressed by
ensemble prediction systems. The ensemble members are obtained by running the NWP model multiple times with small variations in the model and/or in initial and boundary conditions \citep{GneitingRaftery2005, LeutbecherPalmer2008}. The resulting forecast ensemble is probabilistic in nature and allows to quantify forecast uncertainty.
It has become common practice to employ statistical post-processing models to correct for biases and dispersion errors in forecast ensembles, based on training data from
the past. By now, there exists a variety of post-processing models accounting for different needs in the forecasting setting. The most commonly known  models are the so-called \emph{ensemble model output statistics} (EMOS, \citealp{Gneiting&2005}) and
\emph{ensemble Bayesian model averaging} (BMA, \citealp{Raftery&2005}). Both are frequently used techniques that yield full predictive distributions from ensemble output. Overviews on post-processing models can, e.g., be found in \citet{WilksHamill2007},
\citet{Schefzik&2013}, \citet{GneitingKatzfuss2014}, and  \citet{Enspostprocessing2018}.

The original approaches described in \citet{Raftery&2005} and \citet{Gneiting&2005} were developed for temperature and pressure forecasts, where it is reasonable to assume the underlying model distribution to be Gaussian. For other weather variables, however, the Gaussian restriction is often inadequate and alternative distributions are required. For these cases, modifications of the original models have been developed, see, e.g., \citet{Schefzik&2013}, \citet{GneitingKatzfuss2014} or \citet{Hemri&2014} for an overview on models for non-Gaussian weather variables.

We propose a more general post-processing approach, which lets us include non linear relationships between the (observed) weather variable and its respective ensemble forecasts as well as non-Gaussian predictive distributions. For this, we use a copula approach to characterize the joint dependence between the weather variable and the ensemble forecasts.

Gaussian copulas have already been employed
successfully in hydrological \citep{GenestFavre2007, Hemri&2015} and climatological \citep{SchoelzelFriederichs2008} applications. There is also increased interest in using copula based methods to handle dependencies in ensemble post-processing. For example \citet{MoellerLenkTho2013}, \citet{BaranMoeller2015}, \citet{BaranMoeller2016}, and \citet{Schuhen&2012} employ Gaussian copulas for modelling dependencies between weather variables. Furthermore, methods based on the empirical copula enjoy increased popularity as they require little computational effort and allow simultaneous modelling of temporal, spatial and inter-variable dependencies. The most popular approaches are \emph{ensemble copula coupling} (ECC, \citealp{Schefzik&2013, Schefzik2015}) and its modifications \citep{Schefzik2016a, Schefzik2016b}. These methods retain the original multivariate dependence structure present in the raw ensemble, which is often destroyed by applying (univariate) post-processing methods only to a single weather quantity, station location and time point or forecast horizon.
A current overview on the use of copulas in ensemble post-processing can be found in \citet{Enspostprocessing2018}.

We advocate a copula based post-processing approach for a single weather variable at a fixed location and forecast horizon that takes advantage of the dependence structure between the variable and its ensemble forecasts. Typically, post-processing models do not explicitly consider this type of dependency. However, further extensions that combine joint modelling of a weather variable and its forecasts with the above described settings (inter-variable, spatial, temporal models) are possible and part of future work that is discussed in the outlook Section \ref{sec:discussion}.

To allow for maximal dependence flexibility in our new post-processing approach we employ a pair copula construction (PCC) to construct a D-vine copula as discussed in \citet{aas2009pair}. The first such construction was given by \citet{joe1996families} and later organized using a graphical structure called vine by \citet{Bedford2001,Bedford2002} with a focus on Gaussian vines. The development of stepwise estimation \citep{aas2009pair} and selection
\citep{dissmann2013selecting} procedures together with the availability of the software package \texttt{VineCopula}  \citep{VineCopula} has sparked considerable interest among researchers striving to include non-Gaussian dependence in their models. For current developments in this area see \citet{joe2014dependence} and the webpage \texttt{vine-copula.org}.

The novel post-processing method utilizes the D-vine quantile regression approach of \citet{kraus2017d}, implemented in the \texttt{R} package \texttt{vinereg} \citep{VineReg2018}. The approach includes a forward selection of predictors, which are the individual ensemble members in our case. This provides a novel feature in the context of post-processing, as it potentially allows to select informative ensemble members in a data driven manner. This feature can be explored in future research as possible alternative to forming a post-processing model that accounts for groups among ensemble members (see Section \ref{sec:EMOS}). Usually these groups have to be defined manually by the user based on properties of the ensemble. However, the proposed new method offers the possibility to let the model automatically select e.g. group-representative ensemble members.

The paper is organized as follows. The benchmark post-processing approach ensemble model output statistics (EMOS) is reviewed in Section \ref{sec:EMOS}. Section \ref{sec:vine} provides the necessary background on vine copulas and distributions. It also outlines the novel D-vine post-processing method. Verification methods are discussed in Section \ref{sec:verification}. A large verification study involving temperature forecasts of the European Center for Medium Range Weather Forecasts (ECMWF) is provided in Section \ref{sec:application}, demonstrating the superiority of D-vine post-processing especially for larger time horizons.
The paper closes with a discussion and outlook section.

\section{Ensemble model output statistics} \label{sec:EMOS}

The original ensemble model output statistics (EMOS) methodology was introduced
by \citet{Gneiting&2005} and designed for weather variables $Y$ that can be assumed to have a Gaussian predictive distribution, such as temperature and pressure.

In the following, let $Y_t$ be a Gaussian weather variable at time point $t$ at some specified station location. Furthermore, denote by $x_{t1},\ldots, x_{tm}$ the $m$ ensemble forecasts obtained from an NWP model valid for time point $t$, at the specified station. The Gaussian EMOS model combines the $m$ ensemble members as predictors in a multiple linear model
\begin{equation}
\label{emos}
Y_t = a + b_1 x_{t1} + \cdots + b_m x_{tm} + \varepsilon_t, \;\; t=1\ldots,T,
\end{equation}
where $a, b_1, \ldots, b_m$ are real-valued regression coefficients
and $\varepsilon_t \sim \mathcal{N}(0, \sigma^2)$. Given the Gaussian assumption the predictive distribution of the weather variable $Y_t$ is normal with mean
$$\mu_t = a + b_1 x_{t1} + \cdots + b_m x_{tm}.$$

Standard regression theory assumes a homogeneous model for $Y_t$, that is, a constant variance term $\sigma^2$ across all observations, which are the time points $t=1,\ldots,T$ in our application.

However, \citet{Gneiting&2005} found that a constant variance parameter $\sigma^2$ is not accounting for changes in possible over- and underdispersion of the raw ensemble forecasts. A non-homogeneous variance term can be used to exploit the so-called \emph{spread-skill relationship} \citep{WhitakerLoughe, Barker1991},
a connection between the forecast skill of the ensemble and its uncertainty.
\citet{Gneiting&2005} showed that the non-homogeneous EMOS model, also called non-homogeneous regression (NGR) in the post-processing literature, yields a predictive distribution with more accurate dispersion properties than its homogeneous counterpart.

The variance $\sigma^2_t$ in the non-homogeneous EMOS is modeled as a linear function of the (empirical) ensemble variance $S^2_t$ by
\begin{equation}
\label{emosvar}
\sigma^2_t=c + d S^2_t,
\end{equation}
where $c,d$ are non negative parameters.

Here,
\begin{equation*}
S^2_t=\frac{1}{m-1} \sum_{k=1}^{m}(x_{tk}-\bar{x}_t)^2
\end{equation*}
with ensemble mean $\bar{x}_t=\frac{1}{m} \sum_{k=1}^{m}x_{tk}$.

In \eqref{emos} each ensemble member is assigned its own regression coefficient. However, depending on the way the ensemble was generated other approaches might be more reasonable. Many common ensemble prediction system generate the members based on random perturbations of the best set of initial conditions for the NWP model \citep{LeutbecherPalmer2008}. This technique yields statistically indistinguishable (exchangeable) ensemble members. For ensembles where all members belong to a single
exchangeable group it is common to
set $b_1 = \cdots = b_m$, or equivalently, only use the ensemble mean as predictor in the model. In case there are $g=1,\ldots,G$ groups of exchangeable members, the parameters within a group $g$ are typically assumed to be identical.

The resulting general form of a group model can be written as
\begin{equation}  \label{EMOS}
Y_t = a + b_1 x_{t}^{(1)} + b_2 x_{t}^{(2)} + \ldots + b_g x_{t}^{(g)} + \varepsilon_t, \; \; t=1,\ldots,T,
\end{equation}
where $x_{t}^{(g)}$ is defined as the average over all members in group $g$, i.e., $x_{t}^{(g)} = \frac{1}{|g|} \sum_{k \in g} x_{tk}$. The variance term is parameterized to be non constant as well, as defined in \eqref{emosvar}.

The regression parameters $a$, $b_1$, $b_2$, \ldots, $b_m$ (or $b_g$, respectively) and the variance parameters $c$, $d$ need to be estimated based on a set of training data $D_1=\{y_t,x_{t1},\ldots,x_{tm}, t \in T_1\}$. Then, the fitted predictive distribution is used to perform prediction for data points in a separate test set $D_2=\{y_t, x_{t1},\ldots,x_{tm}, t\in T_2\}$. The optimal size of the training set $T_1$ depends on the data and is usually optimized in the application at hand.

A standard approach in post-processing is to employ a so-called rolling training period. Here, a window of fixed length is chosen as training set, e.g. $T_1=30$ days. The parameters are estimated based on this training data and prediction is performed for the time point $T_1+1$. The procedure starts with the very first $T_1$ data points. Then, the training window is shifted one time point ahead, and the whole procedure is performed again.

\citet{Gneiting&2005} compare classical maximum likelihood estimation to optimizing a scoring rule (such as the \emph{continuous ranked probability score}, CRPS, see Section \ref{sec:verification}) and conclude that the latter yields superior predictive performance, specifically when optimizing the CRPS. For the Gaussian distribution, there is a closed-form expression of the CRPS available. The R \citep{R} package \texttt{ensembleMOS} \citep{ensembleMos} provides an implementation of minimum-CRPS parameter estimation for the case of a normally distributed weather variable.

When the data set considers multiple station locations (which is usually the case), there are two prominent approaches to parameter estimation. The first one, called global or regional apporach, estimates only a single set of parameters from all data. The second one, called local approach, estimates a separate set of parameters for each station available in the data, using only the data from the respective station.

\section{D-vine post-processing} \label{sec:vine}

Since the Gaussian assumption might be too restrictive, we develop a more general post-processing approach. We explicitly model the dependence between the weather variable $Y$ and the respective ensemble forecasts $x_1,\ldots,x_m$ with a copula. This allows to design very flexible joint distributions made up with arbitrary choices of the marginal distributions. Therefore, adaptation of our proposed method to weather variables other than Gaussian is much more straightforward than for standard post-processing techniques.

We will first review the concept of a copula, then introduce the class of D-vine copulas and distributions.

\subsection{Copulas}
To fix ideas, assume that we are interested in the joint distribution of $d$ variables $X_1,\ldots,X_d$ with joint distribution function $F$ and marginal distribution functions $F_j, j=1,\ldots,d$. In the case of continuous variables,
\citet{Sklar1959} showed that the joint distribution function can be uniquely represented as
\begin{align}
\label{sklar-cdf}
F(x_1,...,x_d) &= C(F_1(x_1),...,F_d(x_d)),
\end{align}
where the \emph{copula} $C$ is a distribution function on $[0,1]^d$ with uniform margins. The copula characterizes the dependence among the variables. The corresponding equation on the density level can be expressed as
\begin{align}
\label{sklar-pdf}
f(x_1,...,x_d) &= c(f_1(x_1),...,f_d(x_d)) f_1(x_1)\ldots f_d(x_d).
\end{align}
Lower case letters indicate the associated densities of the distributions occuring in \eqref{sklar-cdf}.
For an introduction to copulas see, e.g., \citet{Nelsen2006} and \citet{Joe1997}.

\subsection{D-vine copulas and distributions}
Prominent members of parametric copula models are the elliptical and the Archemedian copulas (see \citet{Joe1997} for precise definitions). These classes are, however, limited in their dependency patterns. The Gaussian copula does not facilitate tail dependence, while the Student t-copula has only symmetric tail dependence. Archimedean copulas only have a single parameter governing the dependence and induce exchangeable dependence structures among pairs of variables.

Thus, there was a need to extend these multivariate copula classes to include more flexibility. This was achieved by using the so-called pair copula construction (PCC) of \citet{bedford2001probability, Bedford2002}. The basic idea behind the PCC is that the joint dependence can be build up by only bivariate building blocks using conditioning.
This construction requires a building plan, which identifies the pairs of variables together with the conditioning set and the associated bivariate copulas.
Stepwise parameter estimation methods have been proposed in \citet{aas2009pair}. For a more detailed introduction to these concepts see \citet{Czado2010,StoeberCzado2012}, \citet{joe2011dependence} and \citet{joe2014dependence}.
The PCC approach allows for a lot of flexibility since each bivariate copula can be selected individually from a wide range of parametric families.

Since there are many ways to build a PCC, a graphical structure called
\emph{regular vine} was introduced by \citet{Bedford2002} to organize the possibilities. A vine consists of a set of nested trees, where each edge of the graph corresponds to a bivariate copula (called \emph{pair copula}).

We restrict ourselves to a subclass of the regular vine copulas called D-vine copulas. In a D-vine, each tree is a path, where the nodes of the current tree are the edges in the subsequent tree. They are easy to construct and conditional distributions of interest are available in closed form.

To shorten notation, we use the abbreviativions $a:b$ for the integer sequence $(a,a+1,\ldots, b)$, and ${\bm x}_{a:b}$ for the sub-vector $(x_a,\ldots,x_b)$. The general form of a D-vine density \citep{Czado2010} is given by
\begin{align}
\begin{aligned} \label{dvineden}
        &\phantom{=}\,f(x_1,...,x_d) \\
        & =  \underbrace{\prod_{j=1}^{d-1} \prod_{i=1}^{d-j}
	c_{i,i+j;(i+1):(i+j-1)}\bigl(u_{i|i+1:(i+j-1)}, u_{i +j| (i+1):(i+j-1)}\bigr)}_{\text{pair copula densities}} \times \underbrace{\prod_{k=1}^{d} f_k(x_k).}_{\text{marginal densities}}
\end{aligned}
\end{align}	
Here $c_{i,i+j;(i+1):(i+j-1)}$ is the bivariate copula density associated with the bivariate distribution $(X_{i},X_{i+j})$ given $X_{i+1}=x_{i+1},...,X_{i+j-1}=x_{i+j-1}$, and
\begin{align*}
u_{i|(i+1):(i+j-1)} &= F_{i|(i+1):(i+j-1)}(x_i|{\bm x_{i+1:i+j-1}}), \\
u_{i + j|(i+1):(i+j-1)} &= F_{i+j|(i+1):(i+j-1)}(x_{i+j}|{\bm x_{i+1:i+j-1}}).
\end{align*}
The conditional distributions in the arguments of $c_{i,i+j;(i+1):(i+j-1)}$ can be computed recursively by using the chain rule of differentiation \citep{joe1996families}.
To allow for easy estimation we make the simplifying assumption \citep{stoeber2013simplified} by assuming that the conditioning values $x_{i+1},\ldots, x_{i+j-1}$ only influence the arguments of the conditional copula density.

\begin{figure}[ht!]%
\centering
\includegraphics[width=9cm, height=5cm]{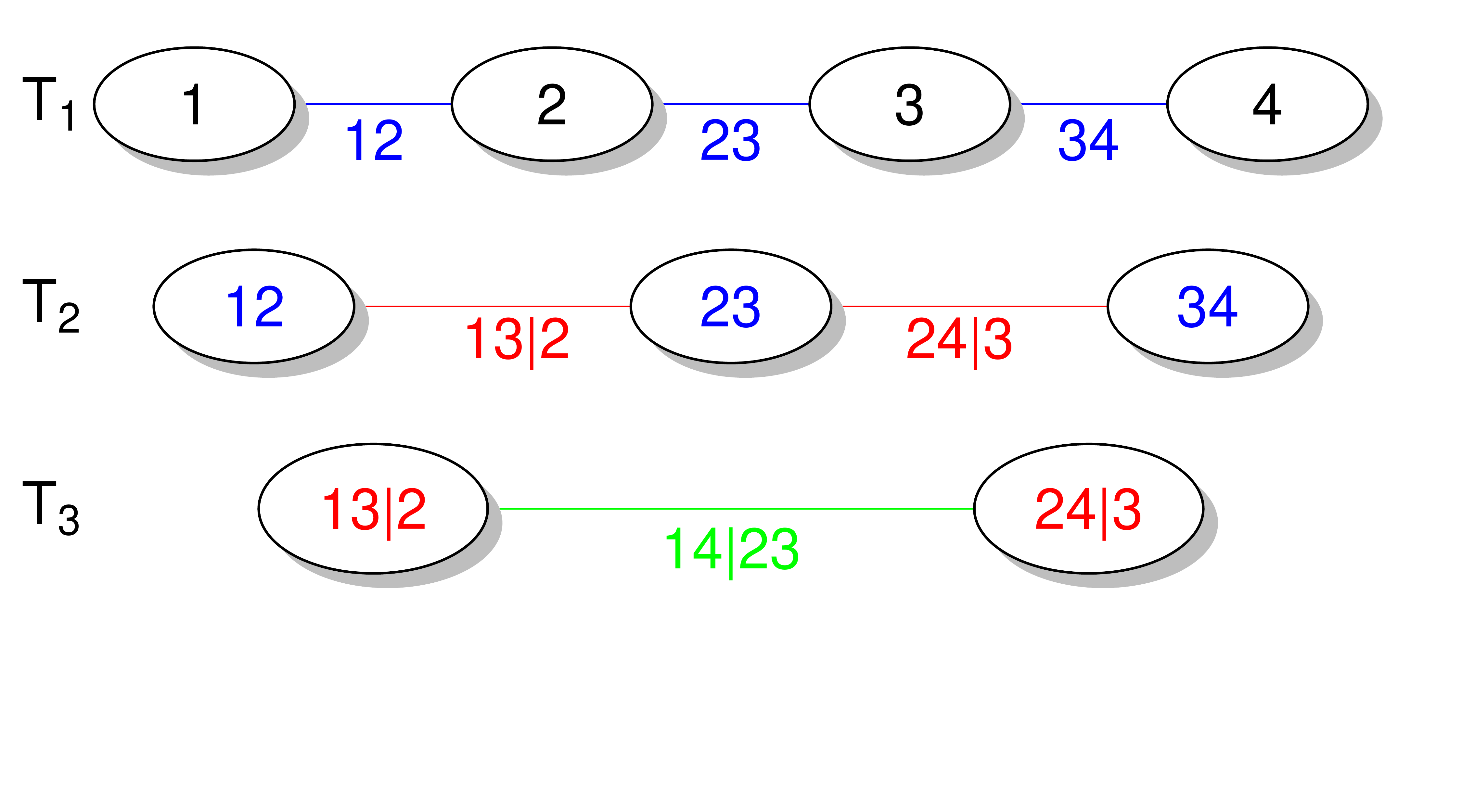}
    \caption{\label{DVineIllustration} Graph theoretic illustration of a D-vine with order $X_1-X_2-X_3-X_4$. The nodes of the trees are plotted
in black circles and the corresponding pair-copulas in colored numbers.}
\end{figure}

To illustrate a D-vine in more detail we outline the case $d=4$ with variables ordered from 1 to 4. In the following example the dependence between variables $X_1$ and $X_2$, $X_2$ and $X_3$, and $X_3$ and $X_4$ are modeled directly, while all other pairwise dependencies are modeled indirectly by conditioning. For $d=4$ the simplified D-vine density has the form
\begin{align*}
f(x_1,x_2,x_3,x_4)  = & \prod_{i=1}^4 f_i (x_i) \times
c_{12}(x_1,x_2) \times c_{23}(x_2,x_3) \times c_{34}(x_3,x_4)  \\
&  \times  c_{13;2}(F_{1|2}(x_1|x_2), F_{3|2}(x_3|x_2)) \\
& \times  c_{24;3}(F_{2|3}(x_2|x_3), F_{4|3}(x_4|x_3) )\\
& \times   c_{14;23} (F_{1|23}(x_1|x_2,x_3),
F_{4|23}(x_4|x_2,x_3)).
\end{align*}
Figure \ref{DVineIllustration} shows a representation of the nested trees defining the D-vine.

The order of the variables determines the complete structure of a D-vine model. This order can be estimated by solving a traveling sales man problem \citep{kruskal1956shortest}, where the weights might be chosen to be absolute values of pairwise Kendall's $\tau$ estimates as was proposed by \citet{dissmann2013selecting} for the more general class of regular vines. D-vine copulas can be fitted using the package \texttt{rvinecopulib} of \citet{rvinecopulib} in \texttt{R} \citep{R}. To do so we need to transform the original data $x_j$ with the probability integral transform (PIT) to the copula data $u_j:=F_j(x_j)$. Since in general the marginal distributions $F_j$ are unknown, we use a univariate kernel density estimator implemented in \texttt{kde1d} of \citet{kde1d}.  This choice  still allows for unique representation of the marginal quantile function, which is needed for the proposed D-vine quantile post-processing approach.

\subsection{Post-processing based on D-vine quantile regression}
We are interested in obtaining a predictive distribution for the weather variable $Y$ to be forecasted using $m$ ensemble forecasts $X_1,\ldots,X_m$. The basic idea is to fit a joint distribution of the weather variable and its forecasts and exploit possible dependencies among them. From the joint distribution we can derive the conditional distribution $Y|X_1,\ldots,X_m$, which represents the post-processed predictive distribution of $Y$ given the $m$ ensemble forecasts.
This goal can be accomplished using quantile regression, i.e. predicting the quantiles of a response $Y$ (weather variable) given a set of predictors $X_{1},\ldots,X_{m}$ (ensemble forecasts).

Predicting conditional quantiles of a random variable
has recently attracted much interest and has applications in various fields. The most frequently used approach is linear quantile regression \citep{koenker1978regression}, an analogue of ordinary least squares for the conditional quantile (instead of the conditional mean).
Quantile regression approaches are also  familiar in the context of ensemble post-processing and weather prediction, specifically when it comes to predict summary statistics of precipitation or wind speed (see, e.g., \citealp{Bremnes2004a, Bremnes2004b, Taillardat&2016, BenBouallegue2017}).

In linear as well as non-linear quantile regression, the issue of quantile crossing may occur. \citet{bernard2015conditional} show that quantile functions in linear regression may cross if non-Gaussian
multivariate data is modeled.
The linear quantile regression method by \citet{koenker1978regression} has been further criticized by \citet{bernard2015conditional} for imposing too restrictive assumptions on the shape of the conditional quantiles.
We utilize a recently proposed extension, the D-vine based quantile regression \citep{kraus2017d}. The method states no explicit assumptions about the shape of the conditional quantiles, and allows the dependence relationship of response and
predictors to be flexibly modelled by a D-vine copula. In this approach a crossing of quantile functions corresponding to different quantile levels is not possible due to model construction.

In the D-vine quantile regression approach, we construct a D-vine that uses $Y$ as the first node in the first tree. In the illustrative example in Figure \ref{DVineIllustration} this means that $Y$ corresponds to node ``1'' in the tree $T_1$. Fixing $Y$ as first node in the first tree results in fitting a D-vine with order $Y-X_{i_1}-...-X_{i_m}$, where $(i_1,\ldots,i_m)$ is an arbitrary permutation of $(1,\ldots,m)$. In this case, it is possible to give a simple expression for the conditional density of $Y$ given the observed values $x_1,\ldots,x_m$, i.e., the post-processed predictive density.

Let $V=F_Y(Y)$ and $U_j=F_j(X_j), j=1,\ldots, m$, denote the random PIT values, while
$v=F_Y(y)$ and $u_j=F_j(x_j), j=1,\ldots, m$ denote the observed PIT values. Now the (conditional) predictive density $f(y|x_1,\ldots,x_m)$ can be written as
\begin{align}
\label{conden}
f(y|x_1,\ldots,x_m)& =  c_{V|U_1,\ldots, U_m}(F_Y(y)|F_1(x_1),\ldots,F_m(x_m)) f_Y(y),
\end{align}
where
\begin{align}
     &c_{V|U_1,\ldots, U_m}(v|u_1,\ldots,u_m) \label{copconden} =
        c_{V|U_1}(v|u_1) \\
     &
     \times \prod_{j=2}^m c_{V, \bm U_{j|1:(j-1)}}\bigl(C_{V|{\bm U}_{1:(j-1)}} (v|{\bm u}_{1:(j-1)}),C_{U_j|{\bm U}_{1:(j-1)}}(u_j|{\bm u}_{1:(j-1)})\bigr), \notag
\end{align}
with index abbreviations as defined before Equation (\ref{dvineden}).

While the D-vine based regression approach fixes the response $Y$ to be the first node in the first tree of the D-vine, the order of the predictor variables $X_1,\ldots,X_m$ is not determined a priori. On the contrary, it is possible to identify informative predictors and order them according to their predictive strength.
\citet{kraus2017d} propose a forward selection approach to select the most important predictors by maximizing an AIC-corrected conditional log-likelihood. The idea is to greedily add predictors which improve the conditional AIC corrected log-likelihood the most until no improvement can be found. Pair copula models can be selected from a catalog of parametric families.

In our application the predictors are the ensemble members and the predictor selection procedure could potentially be utilized to build an analog to a group model. The automatic and data driven selection of the most predictive ensemble members could facilitate an alternative way to manually define groups among the members. This potential feature will be explored in more detail in future research.

After we have fitted the D-vine regression model \eqref{conden}, given the weather variable  observations $y_t$ and the associated  $m$ ensemble forecasts $x_{t1},\ldots,x_{tm}$ for days $t$ in the training set $\{1,\ldots,T\}$ we use the distribution associated with the conditional density  \eqref{conden} as predictive distribution for the test day $Y_{T+1}$ given the ensemble forecasts $x_{T+1,1},\ldots x_{T+1,m}$.

To estimate the CRPS verification metric introduced in the next section, we characterize the conditional predictive distribution in terms of conditional quantiles $F^{-1}_{Y|X_1,\ldots,X_m}(\alpha|x_1,\ldots, x_m)$, evaluated on a dense grid of $\alpha$ levels.
Using the random PIT variables $V=F_Y(Y)$, $U_j=F_j(X_j)$, and corresponding observed PIT values $v=F_Y(y)$, $u_j=F_j(x_j)$, the conditional distribution of $Y$ given the observed values $x_1,\ldots,x_m$ can be expressed as
\begin{align*}
F_{Y|X_1,\ldots,X_k}(y|x_1,\ldots,x_k) &= P(Y \leq y|X_1=x_1,\ldots,X_m=x_m) \\ &= P(F_Y(Y) \leq v|F_1(X_1)=u_1,\ldots,F_m(X_m)=u_m) \\
&= C_{V|U_1,\ldots,U_m}(v|u_1,\ldots,u_m)
\end{align*}

Inversion of this conditional distribution yields the corresponding conditional quantile function at level $\alpha$ as
\begin{align}
\label{condquantile}
 F^{-1}_{Y|X_1,\ldots,X_m}(\alpha|x_1,\ldots, x_m) =  F_Y^{-1}\left(C^{-1}_{V|U_1,\ldots,U_m}(\alpha|u_1,\ldots, u_m)\right)
\end{align}

The required conditional copula distribution function $C_{V|U_1,\ldots,U_m}$ can be obtained as the integral of the conditional copula density $c_{V|U_1,\ldots,U_m}$ given in \eqref{copconden}. \citet{kraus2017d} show that $C_{V|U_1,\ldots,U_m}$ can be computed recursively involving only univariate functions.

An estimate of the conditional quantile function can then be obtained by estimating $F_Y$ and $F_j$, $j=1,\ldots,m$, as well as the conditional copula distribution function  $C_{V|U_1,\ldots,U_m}$ based on integrating \eqref{copconden} and plugging them into Equation (\ref{condquantile}):
$$\hat{F}_Y^{-1}\left(\hat{C}^{-1}_{V|U_1,\ldots,U_m}(\alpha|\hat{u}_1,\ldots, \hat{u}_m)\right).$$

The recursive nature of the computation for the conditional distribution function can be transferred to a recursive algorithm for the quantile function, involving only inverses of univariate functions. These computations are implemented in the
R package \texttt{vinereg} \citep{VineReg2018}.

\section{Verification methods} \label{sec:verification}

This section briefly reviews methods to assess the predictive quality of
probabilistic forecasts.
In the application section calibration of predictive distributions is assessed via rank and
PIT histograms. Furthermore, proper scoring rules are computed as overall performance measures.

The goal of probabilistic forecasting is to maximize sharpness of the predictive distribution subject to calibration. This idea was stated for example in \citet{Gneiting&2005} and \citet{GneitingBalabdaouiRaftery2007}, and picked up by many others. Calibration is a joint property of forecasts and observations, referring to their statistical compatibility. Sharpness  refers to the spread of the predictive distribution and is thus a property of the forecasts only.

To assess the calibration of an ensemble of size $m$, the verification rank histogram (VRH) is typically used. It plots the frequency of the rank of the observation, when pooled with the respective
ensemble members \citep[see, e.g.,][]{Wilks2011}. For the (estimated) predictive distributions, the analogue of the rank histogram is the probability integral transform (PIT) histogram \citep{Dawid1984, GneitingBalabdaouiRaftery2007}. The
PIT is defined as the value $p$ of the predictive distribution function $F$ evaluated at the verifying
observation, that is $p=F(y) \in [0,1]$. In case $y$ would be a sample from $F$, the PIT value $p$ follows a uniform distribution. Therefore, a histogram of PIT values that is close to uniformity indicates a well calibrated forecasting distribution $F$.

Scoring rules for the verification of deterministic or probabilistic forecasts are well known and have been widely used in forecast assessment, as they provide convenient summary measures for forecast quality.

The \emph{continuous ranked probability score} (CRPS,
\citealp{MathesonWinkler1976, Unger1985, Hersbach2000, GneitingRaftery2007, Wilks2011, Gneiting2011}) is a very attractive scoring rule as it is measured in the same units as the observations and provides simultaneous assessment of sharpness and calibration of a predictive distribution.
Suppose that $F$ is a predictive distribution and $y$ is the observed value. The CRPS is defined as
\begin{equation} \label{crps1}
\mathrm{CRPS}(F,y) =\int_{-\infty}^{+\infty} [F(t)-I(y \le t) ]^2 dt
\end{equation}
where $I(\cdot)$ denotes the indicator function on the set $(\cdot)$.

\citet{GneitingRaftery2007} show that if $F$ has a finite first moment, the CRPS \eqref{crps1} can alternatively be written as
\begin{equation} \label{crps2}
\mathrm{CRPS}(F,y)=\mathbb{E}(|Z-y|)-\frac{1}{2}\mathbb{E}(|Z-Z'|),
\end{equation}
where $Z$ and $Z'$ are independent copies of a random variable with distribution $F$.
The CRPS is equal to the absolute error in case of a deterministic forecast.
Equation \eqref{crps2} can be approximated using $R$ deterministic grid points, that is
$$ \mathrm{CRPS}(F,y) \approx \frac{1}{R}\sum_{r=1}^{R} |z_r - y| - \frac{1}{2 R^2}\sum _{r=1}^{R} \sum_{r'=1}^{R} |z_r - z_{r'}|,$$
where $z_r=F^{-1}(\frac{r}{R})$ and $z_{r'}=F^{-1}(\frac{r'}{R})$ for $r,\ r' \in \{1,...,R\}$.
For the D-vine based predictive distribution discussed in the previous section we use the estimated conditional quantiles in \eqref{condquantile} with $\alpha=\frac{r}{R}$ for $r=1,\ldots,R$.

To assess the quality of a probabilistic forecast on a test set of size $T_2$, we average the values of all days in the test set, i.e.
$$\mathrm{CRPS}= \frac{1}{T_2} \sum_{t=1}^{T_2} crps(F_t,y_t)$$
where $y_t \ and\ F_t$ are the observation and the predictive distribution at day $t$, respectively.

For many distributions, including the normal distribution, closed form expressions of the CRPS are available.

\section{Application} \label{sec:application}

In this section we apply our proposed D-vine post-processing to temperature forecasts in Europe and compare its predictive performance to the state-of-the-art EMOS model.
Furthermore, we introduce a novel type of rolling training period that allows to have larger size of the training data while accounting for the time series character of the weather data at the same time.

\subsection{Data} \label{sec:data}

We utilize temperature forecasts from the European Centre for Medium-Range Weather Forecasts (ECMWF, \citealp[see, e.g.,][]{Molteni&1996, Buizza2006, Buizza&2007}). The ECMWF ensemble consists of 50 perturbed members, a control forecast (CTRL) and a high-resolution member (HRES). The 50 members are generated by randomly perturbing initial conditions and model physics to explore the range of uncertainty in the observations and in the model. The control forecast utilizes the most accurate estimate of the current conditions and the currently best description of the model physics. The high-resolution member is obtained by running the NWP model at a higher spatial resolution than the regular members. It is known to improve predictive performance in postprocessing models \citep[see, e.g.,][]{Kann&2009, Gneiting2014, Persson2015}.

Our data set consists of forecasts initialized at 12 UTC for 2-m
surface temperature in Europe along with the verifying observations at 35
different stations in the time period  from 2002-01-01 to 2014-03-20, corresponding to a total number of 4462 forecast cases, see also \citet{Hemri&2014} for an overview on the data.

Furthermore, we consider different forecast horizons of the ensemble, namely 24-h, 48-h, 120-h, and 240-h ahead temperature forecasts.
We perform a separate analysis for each of these forecast horizons to investigate differences in predictive performance for different horizons.

We study five stations in particular, namely Frankfurt-Germany, Madrid-Spain, Bratislava-Slovakia, Gadstrup-Denmark and Vantaa-Finland.

Since the data contains missing values, we perform an imputation based on harmonic regression \citep{Craigmile2011} before applying the post-processing models.
Suppose a time series $Y_t$ contains a periodic component. A model for fitting this component would be
$$Y_t=\mu + a \cos (\psi t) +b  \sin(\psi t) \quad t=1,...,T$$
where $\mu$ is the mean of the time series and $\psi$ is the frequency of periodic variation.
To impute the missing values we fit a linear model
$$Y_t=\mu+ a \cos \left( \frac{2 \pi t}{365.25} \right) +b  \sin \left( \frac{2 \pi t}{365.25} \right), $$
estimating the parameters $a$ and $b$ by least squares.
This procedure was applied separately to each of the forecasts in the ensemble.

\begin{figure}[ht!]%
\centering
\includegraphics[width=10cm, height=6cm]{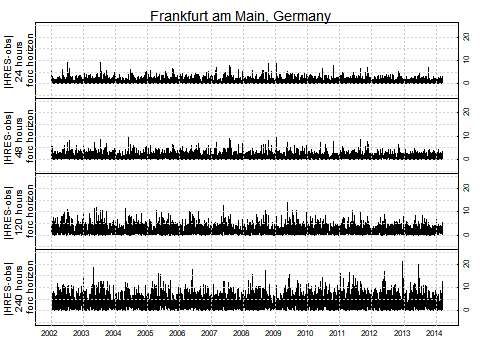}
\caption{\label{absForcError} Absolute differences between observed temperature and respective high-resolution forecasts for different forecast horizons at station Frankfurt.}
\end{figure}

In a preliminary data analysis we investigate the evolution of the forecast error with increasing forecast horizon at the Station Frankfurt (Germany). Figure \ref{absForcError} shows time series of the absolute difference between the observed temperature and the respective forecast made by the high-resolution member for each of the forecast horizons. For larger horizons we clearly see  increasing variability and magnitude in the absolute forecast error. Therefore, we expect that calibrating ensemble forecasts with larger forecast horizons is more difficult.

\subsection{Training period} \label{sec:train}

\subsubsection{Standard version} \label{sec:trainStandard}

When choosing the length of the rolling training period for estimating the parameters of the post-processing model, there is usually a trade off. Short training periods can adapt more quickly to (seasonal) changes, while longer training periods require more data but reduce statistical variability in the estimation. In the post-processing literature, lengths between 20 and 40 days are most common.

For defining a suitable rolling training period of length $T_1$ for EMOS we performed a preliminary analysis.
As we study four different forecast horizons, we conduct the analysis for each of them separately.

We use the last 1000 days as test set for all our methods, that is $T_2=1000$, regardless of the length and type of training period $T_1$. These test days correspond to dates between June 25, 2011 and March 20, 2014. The corresponding $T_1$ days preceding the first test day are used to train the models.

To identify a reasonable training length for the EMOS model, we investigate different lengths for estimating the EMOS parameters, and plot the average CRPS when predicting the 1000 test days with the fitted model as a function of the period lengths.
We performed this analysis for each of the forecast horizons and separately at the five selected stations. Finally, for a fixed forecast horizon, we chose a training length yielding a local minimum of CRPS at all considered stations. The respective choices for $T_1$ based on our preliminary investigation are $T_1=40$ days for 24-h ahead forecasts, $T_1=60$ days for 48-h ahead forecasts, $T_1=100$ days for 120-h ahead forecasts, and $T_1=200$ days for 240-h ahead forecasts.

Intuitively, it seems reasonable to use longer training periods when fitting models that are based on an ensemble with higher forecast horizons.
However, it should be noted that longer training periods (beyond 50 or 60 days) should be considered with care; our data are observed over time with possible autocorrelation structure and seasonal effects.
Therefore, we propose a completely new type of rolling training period in the following Paragraph \ref{sec:trainRefined} to account for these effects.

\subsubsection{Refined version} \label{sec:trainRefined}

For training the D-vine post-processing model we require a much longer training period than for the EMOS model, since many more parameters need to be estimated (all margins and pair copulas). To accomplish the need for a rolling training period of larger size that accounts for the time series character of the data at the same time we propose a novel type of rolling training set that can be applied in case relatively large data sets are available (as in our case study). We call our proposed training period refined rolling training set.

To illustrate how to construct our refined training set, we now consider the prediction of the very first day of our test set, June 25, 2011 based on 24-h ahead ensemble forecasts.
The basic idea is to use the most recent data of the current year preceding the test day, as well as observations of the very same time period but from previous years.
For the current year, we simply take the most recent $n$ observations preceding the test day (as in a standard rolling period). For earlier years, we include an interval of $I=2n$ days that covers the same day of the year as our test day. In general the interval needs not to be symmetric around the test date, however, a symmetric interval is reasonable in the given application. We need to fix the number of days $n$ preceding (and following) the test date. To have a reasonable total length we choose $n=45$ in our application.

To fit the model that predicts June 25 in 2011 our training set then consists of the 45 days before June 25, the 45 days after June 25 as well as June 25 itself in the the nine years preceding 2011, and additionally the 45 days preceding 25 June 2011.

\begin{figure}[ht!]%
\centering
\includegraphics[width=5.5cm, height=5cm]{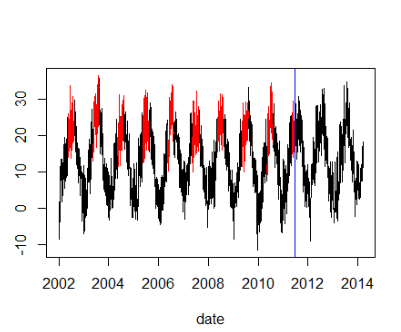}
\includegraphics[width=5.5cm, height=4.3cm]{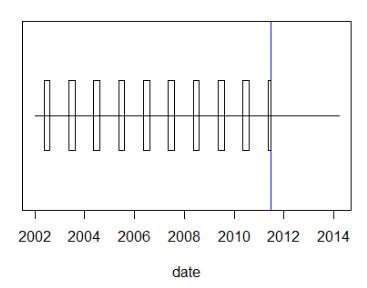}\\
\caption{\label{TrainSetRefined} Illustration of the refined rolling training period when predicting June 25, 2011, based in 24-h ahead forecasts.}
\end{figure}

Figure \ref{TrainSetRefined} illustrates which parts of the full data set are used for the refined rolling training set when predicting June 25, 2011 (marked by a blue vertical line). In the left panel, the observations part of the refined training set are marked in red.

For forecast horizons other than 24-h, only small adaptations are necessary. While the intervals $I=2n$ containing the forecast day in the  years preceding the actual forecast date need not to be changed, an adaptation for the $n$ days before the forecast day in the actual year is necessary.

To further investigate the benefits of the proposed refined rolling training period we also compare estimation of EMOS parameters with the standard training period (with lengths chosen according to our study in Paragraph \ref{sec:trainStandard}) and with the refined training period. We call the two methods EMOS-S and EMOS-R, respectively.

\subsection{Dependence Analysis}

Before we start with the predictive modeling, we conduct a small exploratory analysis. Special attention is paid to whether the Gaussian dependence assumption is valid. We focus on two different stations, Bratislava and Vantaa, and look at 48-h and 240-h forecast horizons using observations from two different dates.

\begin{figure}[ht]
\centering
\includegraphics[width=6cm,height=6cm]{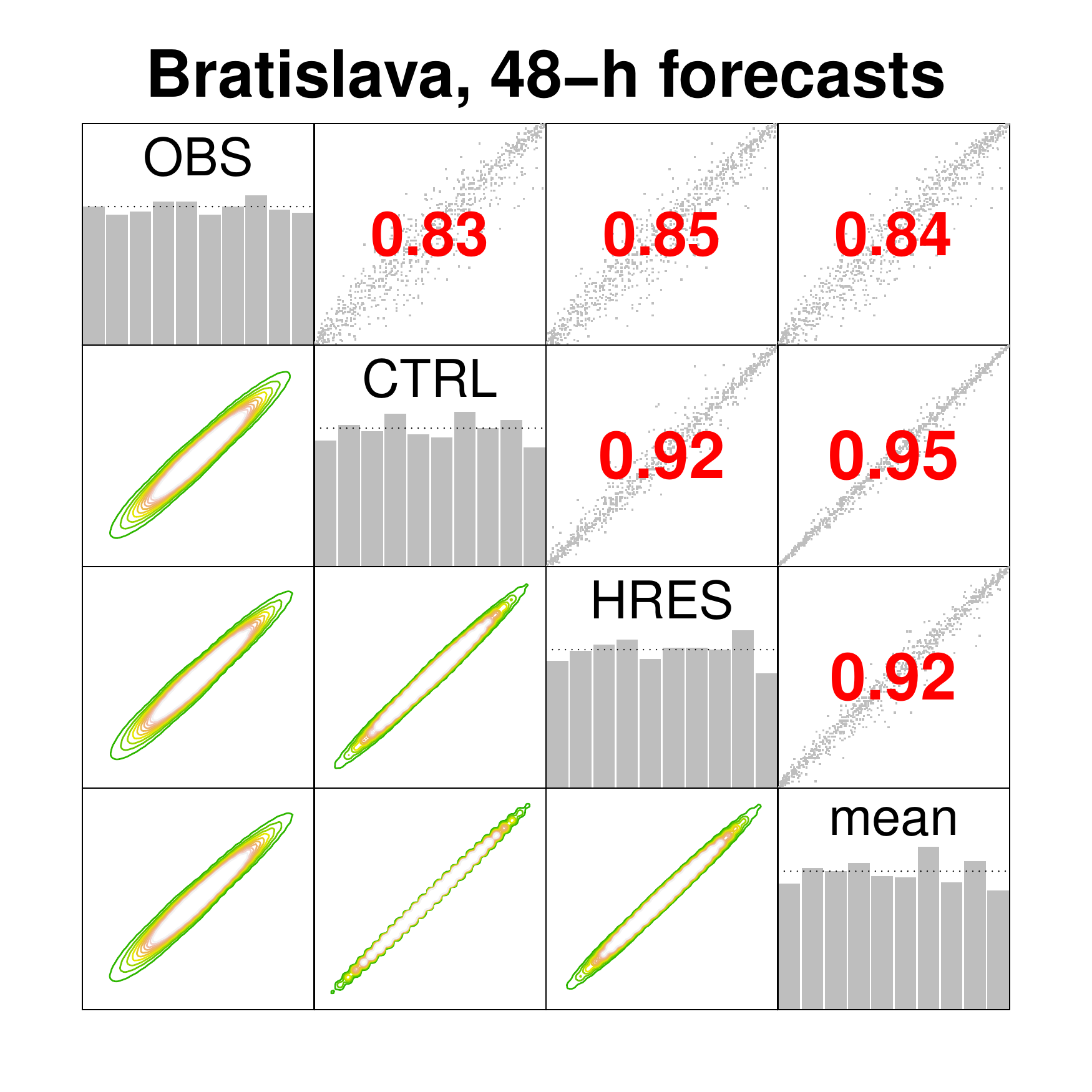}
\includegraphics[width=6cm,height=6cm]{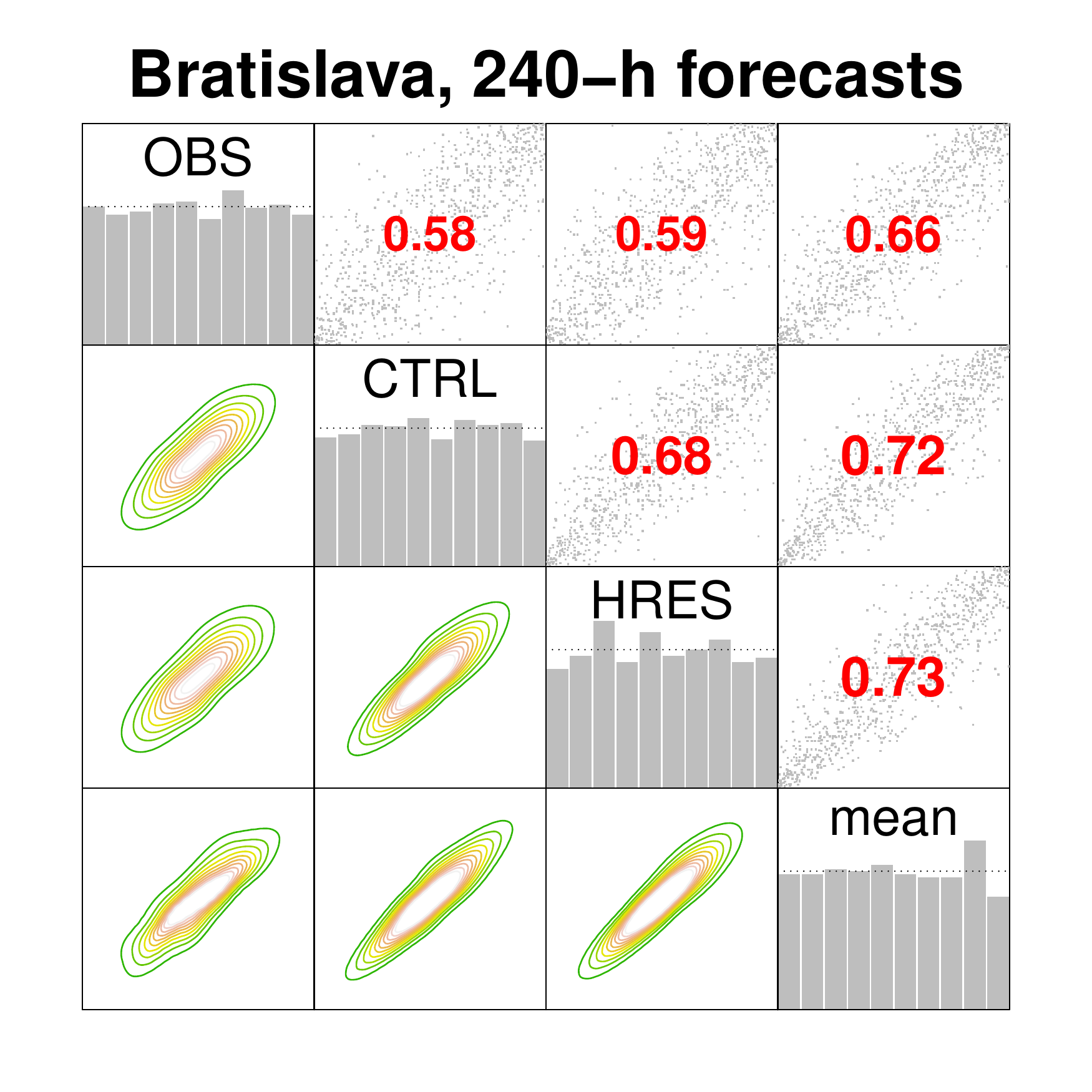}\\
\includegraphics[width=6cm,height=6cm]{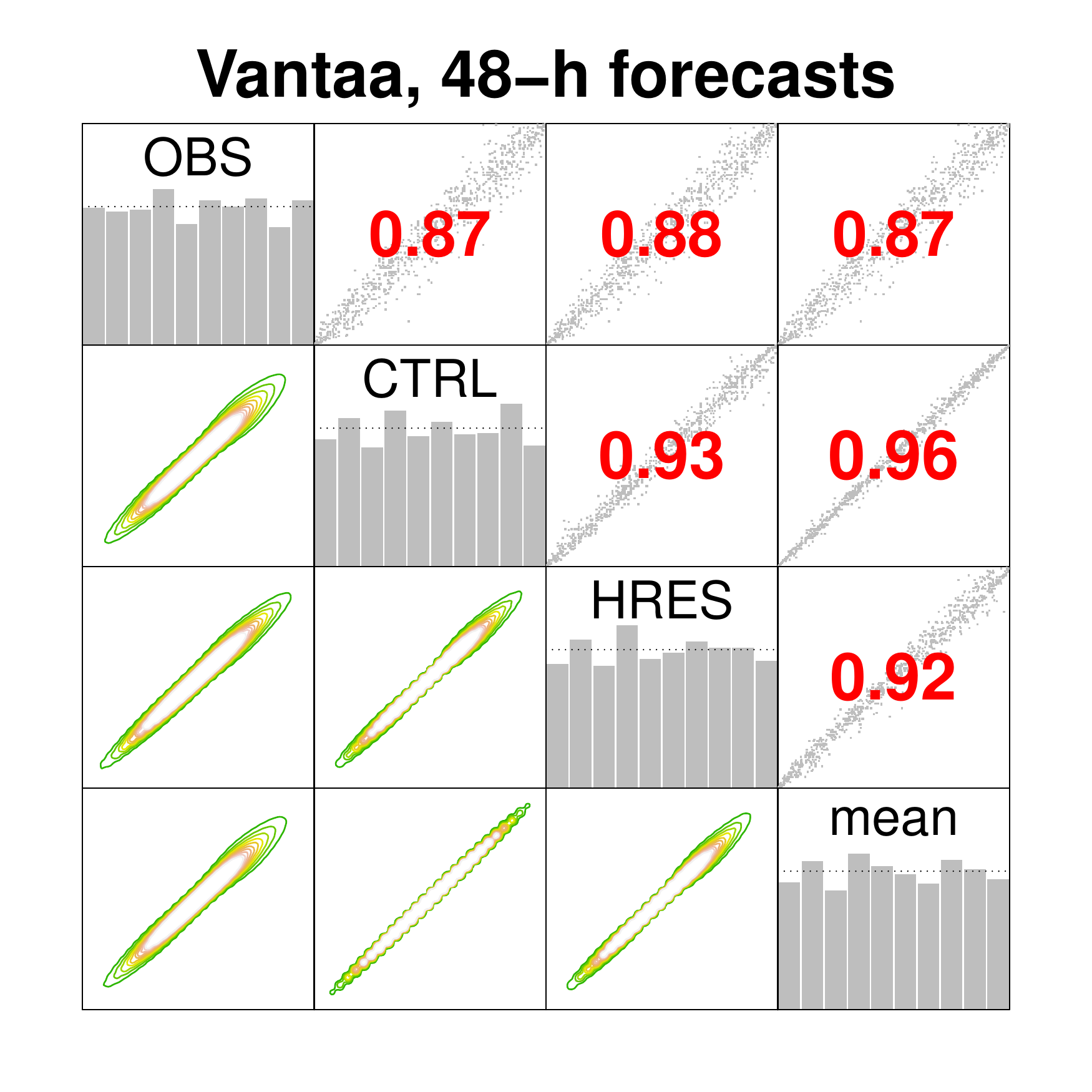}
\includegraphics[width=6cm,height=6cm]{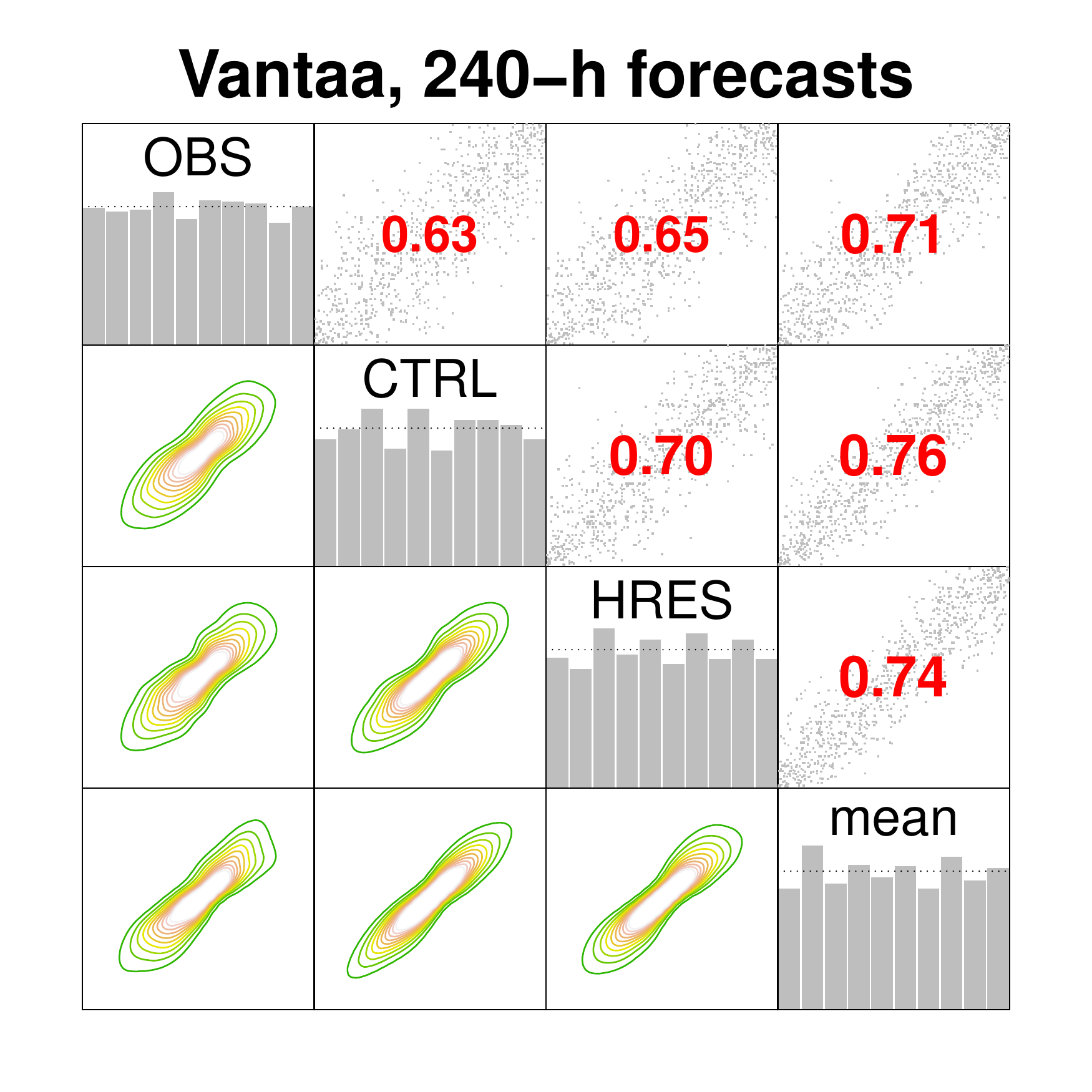}
\caption{Pairwise contour plots for Bratislava and Vantaa based on observations and ensemble forecasts valid on October 16, 2012, for two of the four different forecast horizons.}
\label{DependencePlot1}
\end{figure}

\begin{figure}[ht]
\centering
\includegraphics[width=6cm,height=6cm]{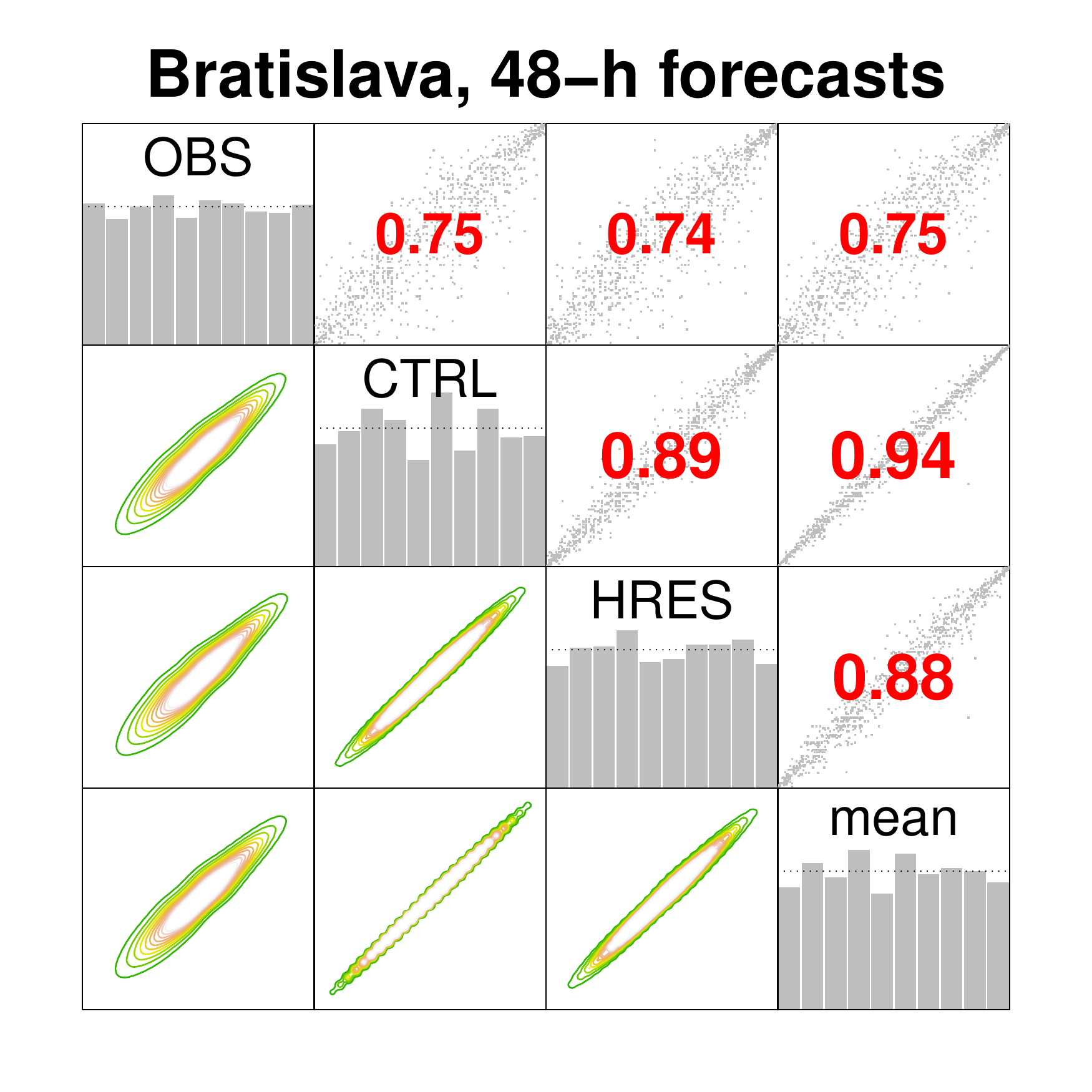}
\includegraphics[width=6cm,height=6cm]{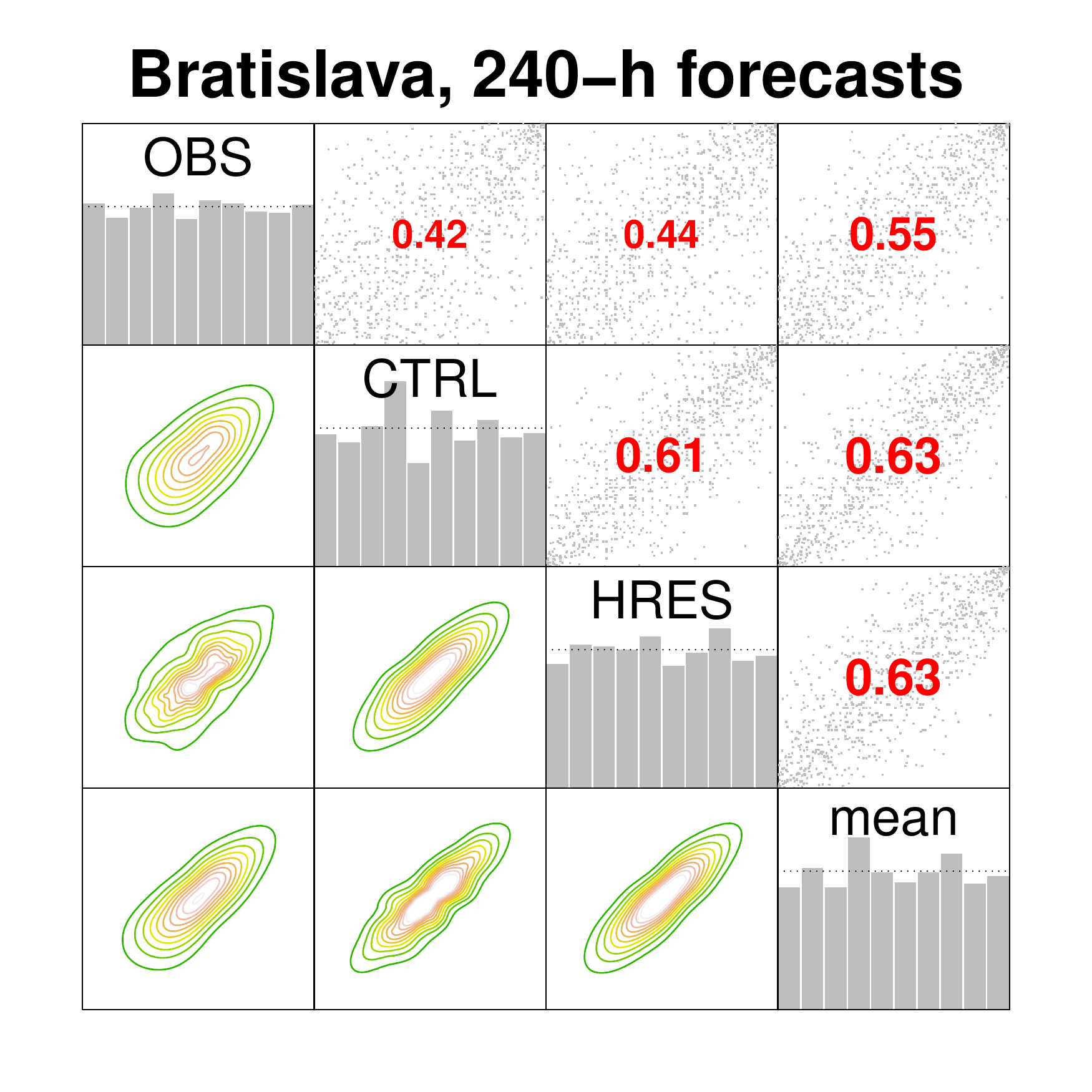}
\includegraphics[width=6cm,height=6cm]{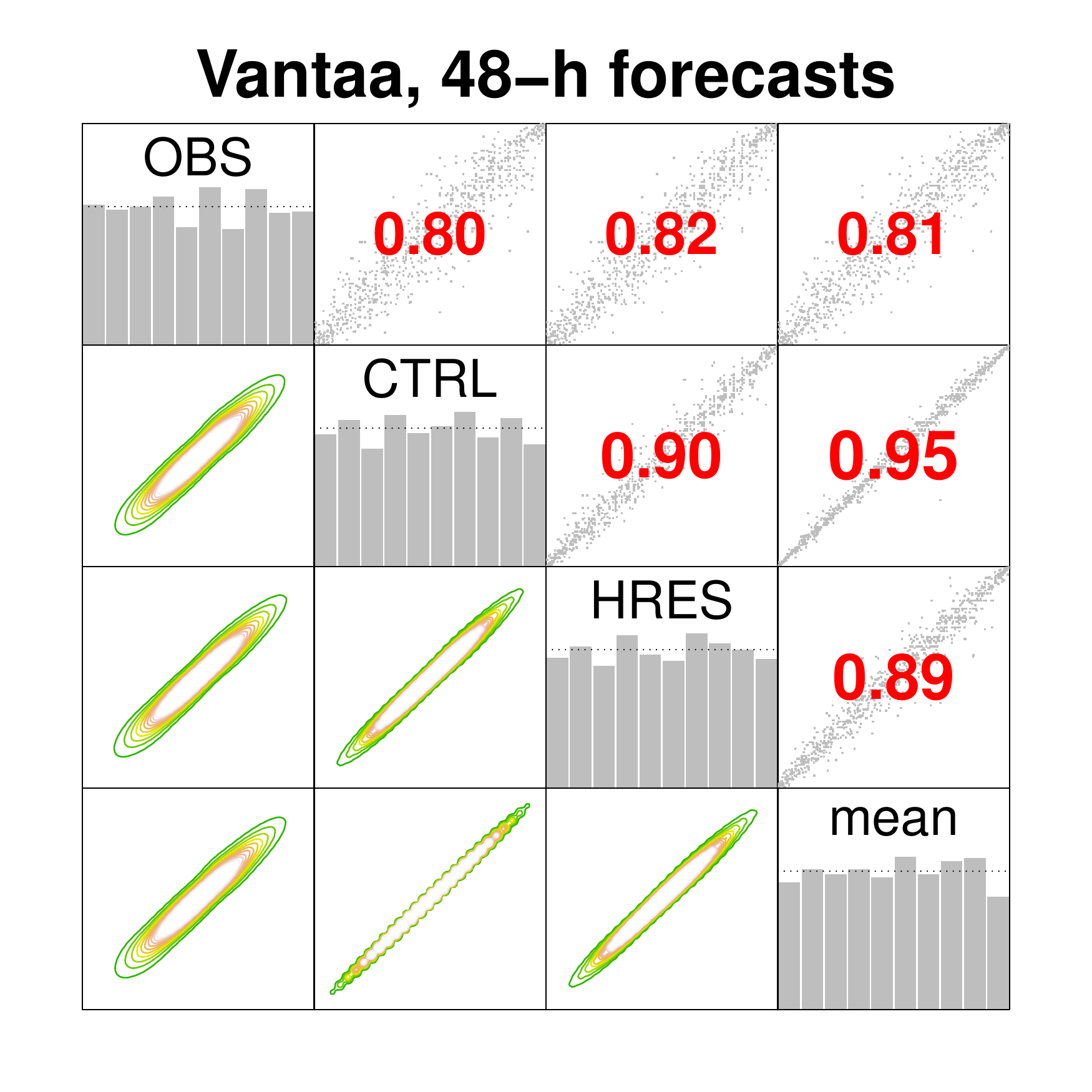}
\includegraphics[width=6cm,height=6cm]{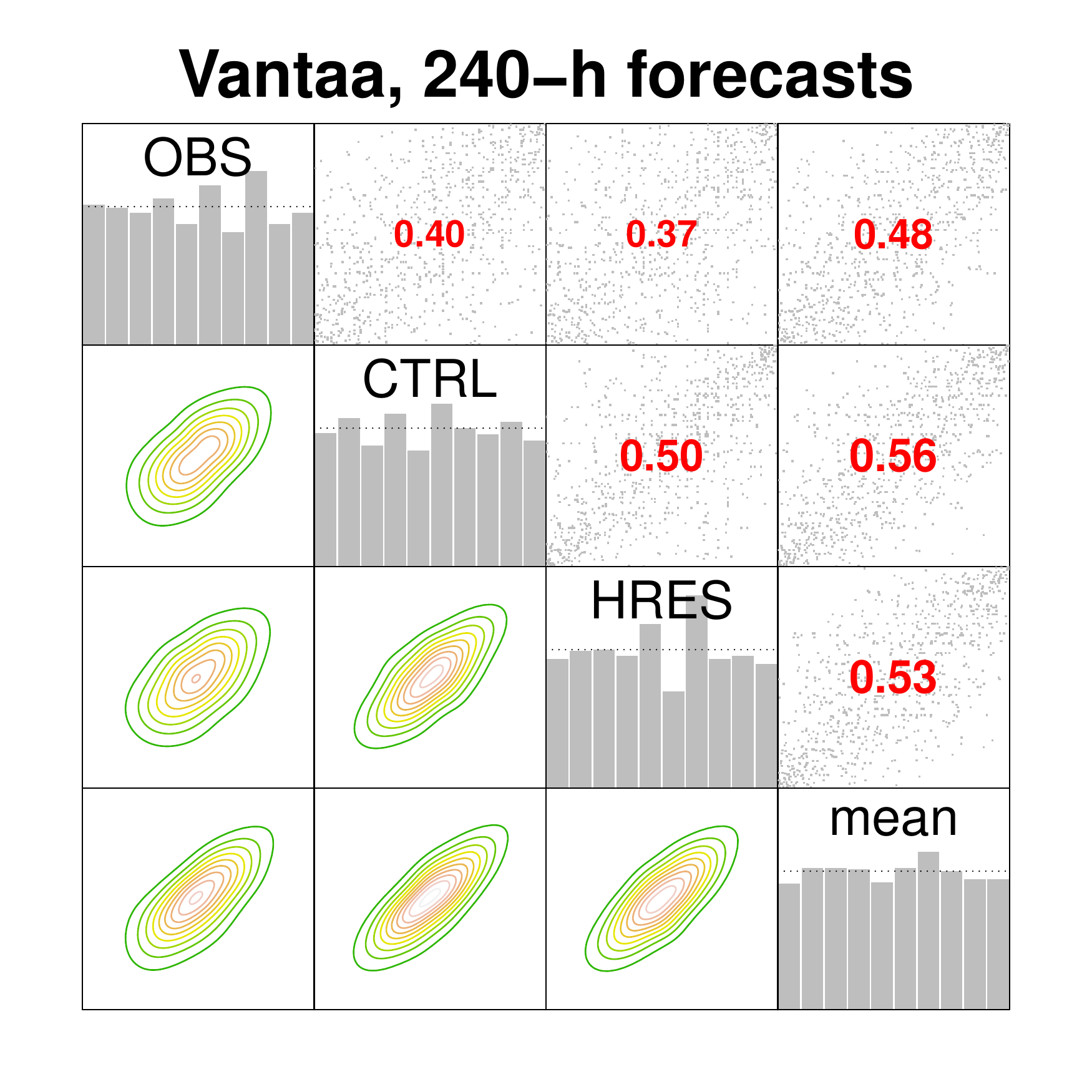}
\caption{Pairwise contour plots for Bratislava and Vantaa based on observations and ensemble forecasts valid on February 8, 2014, for two of the four different forecast horizons.}
\label{DependencePlot2}
\end{figure}

Figures \ref{DependencePlot1} and  \ref{DependencePlot2} summarize the observed dependence for the forecast dates October 16, 2012 and February 8, 2014, respectively. The figures contain information on four variables: the observed temperature (OBS), the control forecast (CTRL), the high-resolution forecast (HRES) and the forecast ensemble average (mean). The diagonal shows PIT-histograms of each variable based on a kernel estimator; the upper right triangle shows pairwise scatter-plots of the PIT observations; the lower left triangle shows corresponding empirical pairwise contour plots. For the latter, a transformation from the restricted domain of the copula data to the unrestricted domain with standard normal margins given by $z_y:=\Phi^{-1}(v)$ and $z_j=\Phi^{-1}(u_j), j=1,\ldots,m$ is made, where $\Phi$ denotes the standard normal distribution function.

Empirical pairwise contours of the transformed data $(z_y,z_1,\ldots,z_m)$ can be used to assess the dependence patterns in the data. In particular, elliptical shapes would indicate Gaussian dependence, which is the basic assumption of the EMOS model \eqref{emos}, while any departures from elliptical shapes point to the need to include non-Gaussian dependencies. In particular, a Student t copula induces star shaped contours, most Archimedean copulas pear shapes and the Frank copula bone shapes (for details see \citealp{StoeberCzado2012}).

Figure \ref{DependencePlot1} shows for both horizons and stations that the dependence structure between observed temperature and forecasts is not Gaussian, since we see non-elliptical shapes. Especially for 240-h the contours of Bratislava are pear shaped (indicating lower tail dependence compatible with a rotated Gumbel copula), while the ones for Vantaa are bone shaped (compatible with a Frank copula).

In Figure \ref{DependencePlot2} we have also non elliptical shapes. For both horizons and stations we see upper tail dependence indicated by the pear shapes.  These shapes indicate that a Gumbel copula  describes the pairwise dependence of the predictive distribution and the forecast members better than a Gaussian copula, thus showing that a non Gaussian prediction model is needed.

Both figures indicate that the dependencies between the predicted temperature and the respective forecasts (and thus their joint distribution) cannot assumed to be Gaussian. Nonetheless, most state-of-the-art post-processing models do not explicitly consider the this dependence, but implicitly assume that the distribution of the response given the forecasts is Gaussian. However, the above results show that it might be worth to explicitly incorporate and model this dependence in a more flexible way. Hence, we can expect the predictive performance in our case study using a D-vine quantile regression to show improvements specifically for higher forecast horizons.

\subsection{Results for predictive performance}

When fitting an EMOS model we employ a group model with $G=3$ exchangeable groups:

\begin{equation}  \label{EMOSCaseStudy}
Y_t = a + b_1 \bar{x}_{t,1:50} + b_2 x_{t,ctrl} + b_3 x_{t,hres} + \varepsilon,
\end{equation}
where $\bar{x}_{t,1:50}=\frac{1}{50} \sum_{k=1}^{50} x_{t,k}$.

Before fitting the final form of the D-Vine model for our application, we performed a preliminary investigation of which predictors (ensemble members) are most frequently chosen in the selection procedure of the D-Vine. For this, we fitted the D-Vine based on all 52 ensemble members plus the average over the 50 perturbed members, thus with 53 predictors in total. This procedure revealed that the high-resolution member, the control member and the average over the 50 perturbed members were the most frequently and earliest selected predictors in the D-vine. The individual 50 perturbed forecasts were chosen less frequently. This result corresponds to the classcial 3-group EMOS model above. However, as already noted, within the D-vine approach the predictor selection is performed automatically based on the data.
For our subsequent analysis, we thus used a D-vine model based on the 3 predictors identified as most important in our preliminary study to speed up computations. When fitting the D-vine based on these 3 predictors, the resulting
selected order was frequently given as $Y-x_{ctrl}-x_{hres}-\bar{x}_{1:50}$.

We now compare the proposed D-vine post-processing to the EMOS method based on the standard rolling training period (EMOS-S), and to the EMOS method based on our refined training period (EMOS-R).
Table \ref{tab:dvineEMOS} shows the CRPS values of the EMOS-S, EMOS-R, and the D-vine model averaged over the 1000 test days for each of the five considered stations.
Bold values indicate the smallest CRPS value within a setting (specific station and forecast horizon).

\renewcommand{\arraystretch}{0.93}
\setlength{\tabcolsep}{1.3mm}
\begin{table}[ht!]
\centering
\begin{tabular}{lrrrrrrr}
  \hline
 Model & Horizon & Training & Frankfurt & Madrid & Gadstrup & Bratislava & Vantaa \\
  \hline
  D-vine & 24h & 864 & 0.8328 & 0.7595 & 0.6026 & \textbf{0.9235} & 0.7179 \\
  EMOS-S & 24h & 40 & \textbf{0.7841} & \textbf{0.6583} & 0.5938 & 0.9645 & \textbf{0.7100} \\
  EMOS-R & 24h & 864 & 0.8034 & 0.7643 & \textbf{0.5899} & 0.9280 & 0.7137 \\
  \hline
  D-vine & 48h  & 864 & 0.9124 & 0.8274 & \textbf{0.6859} & \textbf{1.0316} & \textbf{0.8203} \\
  EMOS-S & 48h & 60 & \textbf{0.8747} & \textbf{0.7316} & 0.7028 & 1.0789 & 0.8359 \\
  EMOS-R & 48h & 864 & 0.8801 & 0.8264 & 0.6881 & 1.0444 & 0.8277 \\
  \hline
  D-vine & 120h & 864 & 1.2444 & 1.1568 & 0.9993 & \textbf{1.4531} & \textbf{1.2186} \\
  EMOS-S & 120h & 100 & 1.2815 & \textbf{1.1238} & 1.0311 & 1.5320 & 1.2438 \\
  EMOS-R & 120h & 864 & \textbf{1.2276} & 1.1456 & \textbf{0.9829} & 1.4558 & 1.2306 \\
  \hline
  D-vine & 240h & 864 & 2.0554 & \textbf{1.7255} & 1.5047 & 2.2935 & \textbf{1.8615} \\
  EMOS-S & 240h & 200 & 2.1248 & 1.7667 & 1.5456 & 2.3969 & 1.8604 \\
  EMOS-R & 240h & 864 & \textbf{2.0419} & 1.7508 & \textbf{1.4910} & \textbf{2.2473} & 1.8694 \\
  \hline
\end{tabular}
\caption{Average CRPS over all test days for D-vine model estimated on refined training period, EMOS model estimated with standard training period (EMOS-S) and refined training period (EMOS-R) for the different forecast horizons at the five selected stations.}
\label{tab:dvineEMOS}
\end{table}

The results indicate that for higher forecast horizons the D-vine model or EMOS based on the new training period exhibit superior performance, while the performance of the standard state-of-the-art EMOS deteriorates. For 48-h ahead forecasts, the D-vine model is superior to the other models at three out of five stations.
At the station Vantaa the D-vine shows superior performance for all higher forecast horizons.

\begin{figure}[ht!]%
\centering
\includegraphics[width=0.99\textwidth]{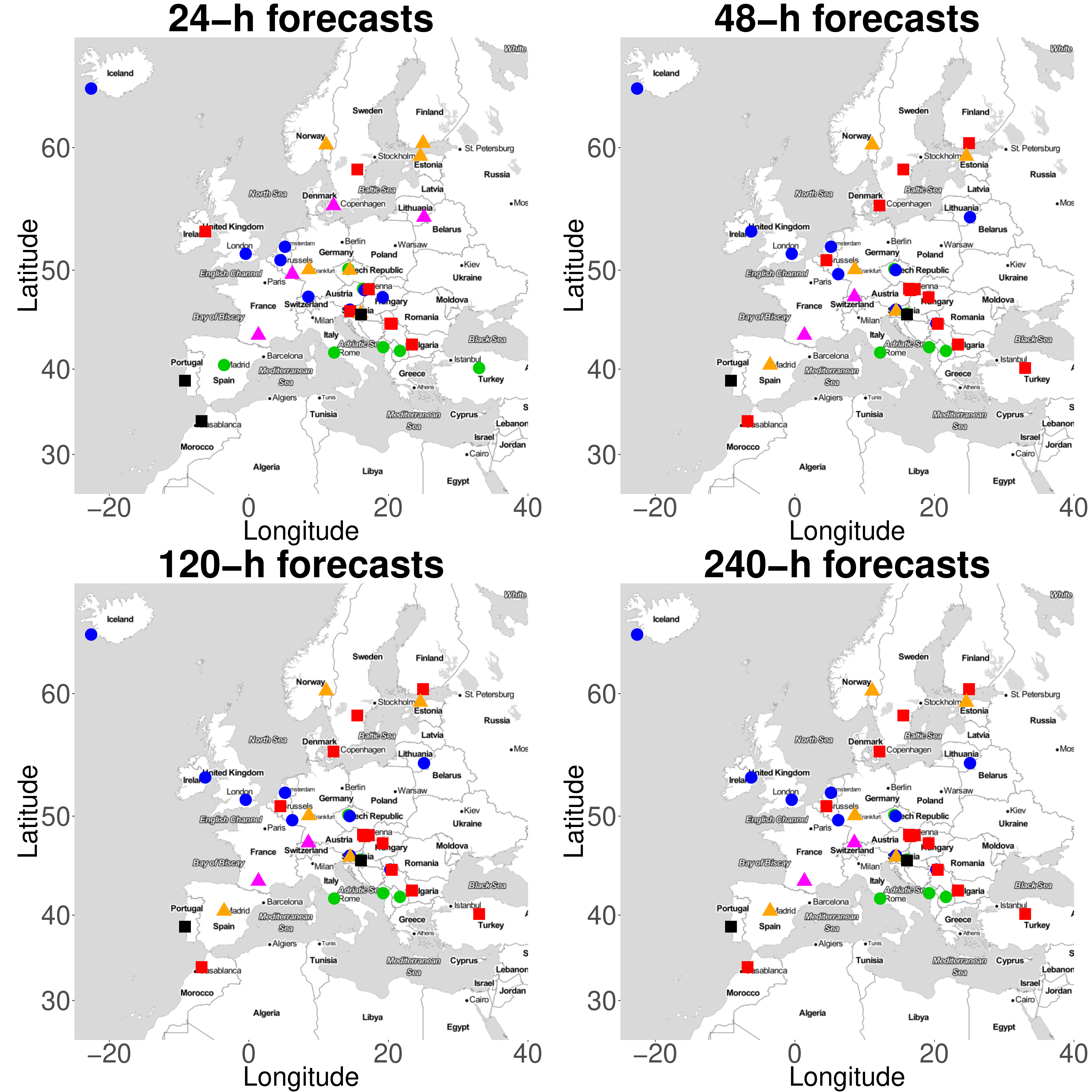}
\caption{\label{StationsCRPS} Map indicating at each of the 35 stations which model has superior performance in terms of CRPS, for each of the forecast horizons. Black square: D-vine $<$ EMOS-S $<$ EMOS-R, red square: D-vine $<$ EMOS-R $<$ EMOS-S, green dot: EMOS-S $<$ D-vine $<$ EMOS-R, blue dot: EMOS-R $<$ D-vine $<$ EMOS-S, orange triangle: EMOS-S $<$ EMOS-R $<$ D-vine, pink triangle: EMOS-R $<$ EMOS-S $<$ D-vine.}
\end{figure}

While the above table shows the results at five selected stations in particular, we now look at the performance at all stations considered in the case study. Figure \ref{StationsCRPS} presents maps with the 35 stations, one map for each of the four forecast horizons considered. At each station a specific symbol-color combination corresponds to one of six rank possibilities among the 3 methods.
Here, a black square indicates the CRPS rank sequence D-vine $<$ EMOS-S $<$ EMOS-R, a red square the sequence D-vine $<$ EMOS-R $<$ EMOS-S, a green dot the sequence EMOS-S $<$ D-vine $<$ EMOS-R, a blue dot the sequence EMOS-R $<$ D-vine $<$ EMOS-S, an orange triangle the sequence EMOS-S $<$ EMOS-R $<$ D-vine, and a pink triangle the sequence EMOS-R $<$ EMOS-S $<$ D-vine. That is, the incidences where the D-vine performs best are represented by squares, the incidences where the D-vine performs second best by dots, and the incidences where the D-vine shows the worst performance of all three methods by triangles.

For 48-h ahead forecasts the D-vine model works particularly well. While the situation is quite mixed for 24-h forecasts, the number of stations where the D-vine model performs best and the refined EMOS second best increases for 48-h forecasts. A similar pattern can be observed in the number of stations where D-vine is second best and refined EMOS best.

When going to even higher forecast horizons the refined training period becomes more and more useful. Here, for most of the stations EMOS with the refined period or the D-vine (also based on refined period) perform best, while at only very few stations the standard EMOS method is superior.

\begin{table}[ht!]
\centering
\begin{tabular}{lccc}
  \hline
   & Square & Dot & Triangle  \\
  \hline
  24h & 0.29 & 0.43 & 0.28 \\
  48h & 0.40 & 0.40 & 0.20 \\
  120h & 0.26 & 0.51 & 0.17 \\
  240h & 0.09 & 0.80 & 0.11 \\
   \hline
   \end{tabular}
\caption{Relative frequency of stations where D-Vine performs best (squares), second best (dots), and worst (triangles) for each forecast horizon.}
\label{FreqRankModels}
\end{table}

Table \ref{FreqRankModels} summarizes the results from Figure \ref{StationsCRPS}, it shows the relative frequency of cases where the D-Vine model performs best (represented by squares in the maps), second best (represented by dots in the maps), and worst (triangles in the maps), for each of the forecast horizons. From the table we see again, that for all forecast horizons the D-vine performs best or second best in the majority of cases.

\begin{figure}[ht!]%
\centering
\includegraphics[width=0.95\textwidth]{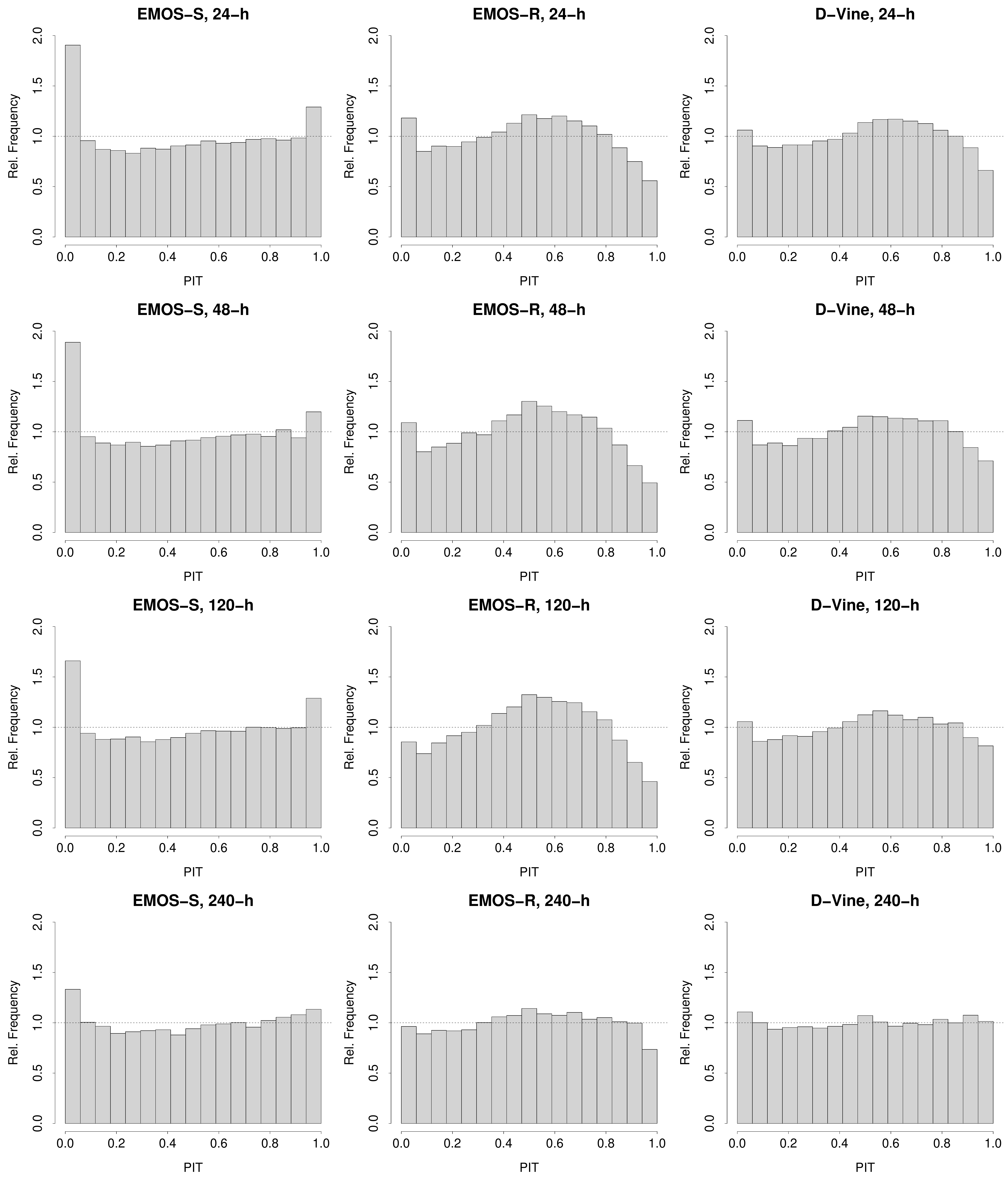}
\caption{\label{PitHists} PIT histograms for the different forecast horizons obtained by the three methods accumlated over all 35 stations.}
\end{figure}

Figure \ref{PitHists} presents the PIT histograms of the three competing models for each of the forecast horizons, accumulated over the 35 stations.

For all forecast horizons we see that the standard EMOS methods exhibits a clearly visible U-shape (indicating underdispersion), which is less pronounced for higher forecast horizons.

When EMOS is fitted station-wise (local EMOS, the approach we use here), the PIT histogram often exhibits signs of underdispersion. This is due to the fact that only data of a specific station is used, while information from other stations is not taken into account. This often leads to deteriorated calibration properties.
On the contrary, EMOS estimated with the refined training period exhibits the reverse effect, the PIT histograms show a more or less pronounced hump shape, an indication for overdispersion.

The D-vine model shows the best PIT histograms for all forecast horizons, it is always closest to uniformity. For 24-h and 48-h ahead forecasts, the D-vine PIT histograms show weak signs of overdispersion, but less pronounced than EMOS-R. For 120-h and 240-h ahead forecasts, the histograms are almost uniform, indicating a high level of calibration.
To conclude, the PIT histograms indicate that the D-vine model improves calibration compared to EMOS-S and EMOS-R, especially for higher forecast horizons.

\section{Discussion and outlook} \label{sec:discussion}

This article presents a novel statistical post-processing approach for NWP forecasts of temperature, which is typically assumed to follow a Gaussian distribution.
Classical post-processing models as Ensemble Model Output Statistics (EMOS) do not explicitly investigate the actual dependence between the observation and its forecasts. In most cases the distribution of the observation given the forecasts is implicitly assumed to be Gaussian as well.

We introduce a D-vine based post-processing approach which explicitly models this dependence. Furthermore, the type of dependence structure in our approach is very flexible due to the pair copula approach and is chosen data driven. Important predictors are identified sequentially, when constructing the D-vine.
This feature could potentially be utilized as an alternative approach to classical group models in post-processing. It is planned to investigate the performance of this approach compared to manually built group models in future research.

In addition to the D-Vine based post-processing model, we propose a new type of (rolling) training period period that is able to use more data for training a post-processing model. Instead of simply using a longer time window, the new period combines a time window of fixed length from several years to accommodate seasonal effects.
Our case study shows that the D-vine model as well as the new training period improve predictive performance specifically for higher forecast horizons, where the joint dependence of observation and forecasts obviously departs from Gaussian.

As the D-vine approach is very general and flexible, it will be straightforward to extend it to the setting of other weather variables. Since the D-vine quantile regression model estimates the dependence structure and type of pair copula automatically in a data-driven way, there is no need to define a modified statistical model suitable for other than Gaussian distributions. For example, the model can easily accommodate asymmetric and heavy-tailed marginal distributions as is often required to model wind speed. Also, even more complex relationships can be captured using nonparametric estimators for the pair copula densities \citep[see, e.g.,][]{nagler2016evading, Nagler2017}.
In future research we will also investigate how spatial or temporal dependencies can be incorporated into the D-vine post-processing method. Certain weather conditions are often persistent over small time periods and spatial neighborhoods. Exploiting these dependencies with an extension of the D-vine model might be highly beneficial.

\section*{Acknowledgements}

Financial support was provided by an award from the program ``Global Challenges for Women in Math Science'' of the Technical University of Munich.
We are also grateful to the European Centre for Medium-Range
Weather Forecasts (ECMWF) and the German Weather Service
(DWD) for providing forecast and observation data, respectively.

\end{document}